\documentclass[12pt,a4paper]{article}
\usepackage{amsmath,amsthm,amsfonts,amssymb,bbm}
\usepackage{graphicx,psfrag,subfigure}
\usepackage{cite}

\numberwithin{equation}{section}

\newcommand{\Pb}{\mathbbm{P}}
\newcommand{\E}{\mathbbm{E}}
\newcommand{\Id}{\mathbbm{1}}
\newcommand{\e}{\varepsilon}
\newcommand{\I}{{\rm i}}
\newcommand{\R}{\mathbb{R}}
\newcommand{\N}{\mathbb{N}}
\newcommand{\Z}{\mathbb{Z}}
\newcommand{\dx}{\mathrm{d}}
\newcommand{\cte}{\mathrm{const\,}}
\newcommand{\Tr}{\mathrm{Tr}}

\newtheorem{prop}{Proposition}
\newtheorem{thm}[prop]{Theorem}
\newtheorem{claim}[prop]{Claim}
\newtheorem{lem}[prop]{Lemma}

\newtheorem{cor}[prop]{Corollary}

\newenvironment{proofOF}[2]{\removelastskip\vspace{6pt}\noindent {\it Proof of #1.}~\rm#2}{\qed \par\vspace{6pt}}

\title{Slow decorrelations in KPZ growth}
\author{Patrik L. Ferrari\thanks{Weierstrass Institute, WIAS Berlin, e-mail: \texttt{ferrari@wias-berlin.de}}}

\date{30. June 2008}

\begin{document}
\maketitle \sloppy

\begin{abstract}
For stochastic growth models in the Kardar-Parisi-Zhang (KPZ) class in $1+1$ dimensions, fluctuations grow as $t^{1/3}$ during time $t$ and the correlation length at a fixed time scales as $t^{2/3}$. In this note we discuss the scale of time correlations. For a representant of the KPZ class, the polynuclear growth model, we show that the space-time is non-trivially fibred, having slow directions with decorrelation exponent equal to $1$ instead of the usual $2/3$. These directions are the characteristic curves of the PDE associated to the surface's slope. As a consequence, previously proven results for space-like paths will hold in the whole space-time except along the slow curves.
\end{abstract}

\section{Introduction}
The well known KPZ equation for the evolution of a non-linear stochastic growth of an interface on a one-dimensional substrate, $x\mapsto h(x,t)$ ($x\in \R$ the space and $t\in\R$ the time), is~\cite{KPZ86}
\begin{equation}\label{eqKPZequation}
\frac{\partial h(x,t)}{\partial t}=\nu\Delta h(x,t)+\tfrac12\lambda (\nabla h(x,t))^2+\eta(x,t).
\end{equation}
$\nu$ reflects the smoothing mechanism due to the surface tension, $\lambda\neq 0$ in front of the non-linear term is responsible for lateral spread of the surface and irreversibility, and $\eta$ is a local noise term. (\ref{eqKPZequation}) corresponds to a situation with small slope $\textbf{u}=\nabla h$. Generically, one has $v(\textbf{u})$ instead of the non-linear term, with $v$ the speed of growth as a function of the slope $\textbf{u}$, having the property $v''(\textbf{u})\neq 0$.

The long time evolution has a deterministic macroscopic behavior, i.e., the limit shape
\begin{equation}\label{eqLimitShape}
h_{\rm ma}(\xi)=\lim_{t\to\infty} \frac{h(\xi t,t)}{t}
\end{equation}
is non-random. To see a non-trivial fluctuation behavior one has to focus on an appropriate mesoscopic scale.

For the KPZ class in $1+1$ dimensions, the \emph{correlation exponent} is $2/3$ and the \emph{fluctuation exponent} is $1/3$~\cite{FNS77,BKS85,BS95}, that is, the rescaled height function at time $t$
\begin{equation}\label{eqRescaledHeight}
h^{\rm resc}(u,t):=\frac{h(\xi t+ u t^{2/3},t)-t h_{\rm ma}(\xi+u t^{-1/3})}{t^{1/3}}
\end{equation}
will converge to a non-trivial process as $t\to\infty$.

The rescaling (\ref{eqRescaledHeight}) makes sense as soon as the macroscopic shape is smooth in a neighborhood of $\xi$. However, depending on the initial conditions, (\ref{eqKPZequation}) can produce spikes in the macroscopic shape. If one looks at the surface gradient $\textbf{u}=\nabla h$, the spikes of $h$ corresponds to shocks in $\textbf{u}$, and it is known that the shock position fluctuates on a scale $t^{1/2}$. For particular models, properties of the shocks have been analyzed, but mostly for stationary growth (see~\cite{Fer90,DJLS93} and references therein). In this paper we discuss spike-free regions of the surface, where the limit shape $h_{\rm ma}$ is smooth.

\subsubsection*{Limit processes at fixed time}
\begin{figure}[t!]
\begin{center}
\psfrag{x}{$x$}
\psfrag{h}{$h(x,t)$}
\includegraphics[height=2.5cm]{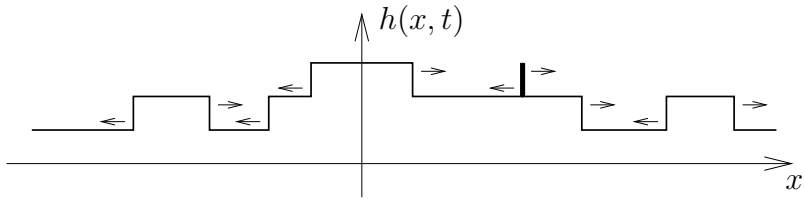}
\caption{Illustration of the PNG height and its dynamics. The bold vertical piece is a nucleation. The arrows indicate the movements of the steps. A Java animation of the PNG dynamics is available at~\cite{FerPNG}.}
\label{FigPNG}
\end{center}
\end{figure}
Recent results obtained by analyzing solvable models in the KPZ class, show that the limit processes are not as universal as the scaling exponent, but one has to divide into a \emph{few} subclasses. One of the most studied model is the polynuclear growth (PNG) model, which we shortly introduce. At time $t$, the surface is described by an integer-valued height function $x\mapsto h(x,t)\in \Z$, $x\in \R,t\in \R_+$, with steps of size one. It thus consists of up-steps ($\lrcorner\hspace{-0.15em}\ulcorner$) and down-steps ($\urcorner\hspace{-0.15em}\llcorner$), see Figure~\ref{FigPNG}. The dynamics has a deterministic and a stochastic part:
\begin{itemize}\label{PNGdynamics}
\item[(i)] up- (down-) steps move to the left (right) with unit speed and disappear upon colliding,
\item[(ii)] pairs of up- and down- steps (nucleations) are randomly added on the surface with some given intensity. The up- and down-steps of the nucleations then spread out with unit speed according to (i).
\end{itemize}
Analytic results have been obtained in the following cases starting with a flat initial substrate $h(x,0)=0$, $x\in\R$.\\[0.5em]
\emph{(a) PNG droplet:} the nucleations form a Poisson point process in space-time with intensity $\rho(x,t)=2$ for $|x|\leq t$ and $\rho(x,t)=0$ otherwise. The limit shape is then $h_{\rm ma}(\xi)=2\sqrt{[1-\xi^2]_+}$.\\[0.5em]
\emph{(b) flat PNG:} the nucleation intensity is $\rho(x,t)=2$ (or any other constant), then the limit shape is simply $h_{\rm ma}(\xi)=2$.
\begin{figure}[t!]
  \begin{center}\hfill
  \subfigure[]{
  \includegraphics[width=6cm,height=4cm]{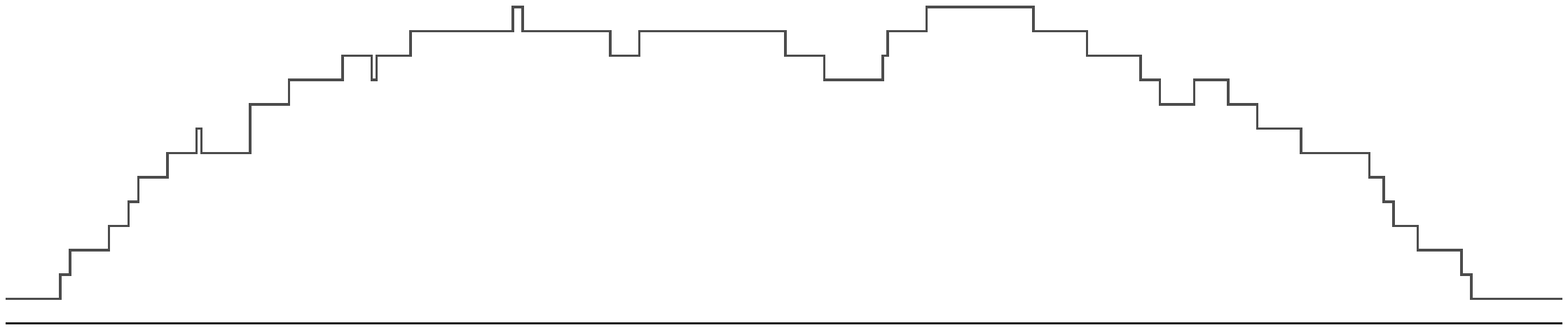}}\hfill
  \subfigure[]{
  \includegraphics[width=6cm,height=3.5cm]{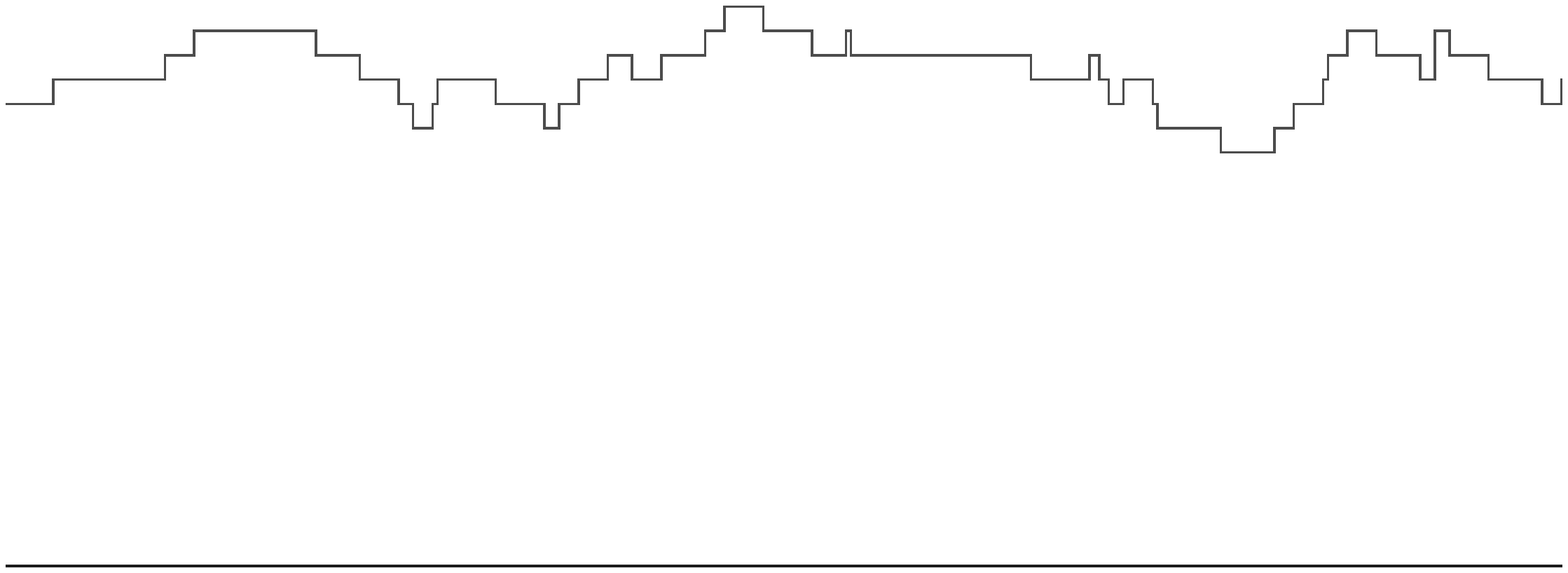}} \hfill \phantom{}
  \caption{Snapshots of the height function for (a) the PNG droplet and (b) the flat PNG geometries.}\label{FigureSnapshots}
\end{center}
\end{figure}

These are the two typical cases obtained starting with non-random smooth initial condition, and, by universality, one expects the following to holds.\\[0.5em]
\emph{(a) Curved macroscopic limit shape} ($h_{\rm ma}''\neq 0$ in a neighborhood of $\xi$). The limit process describing the rescaled surface $h^{\rm resc}$ is the Airy$_2$ process, ${\cal A}_2$,
\begin{equation}\label{eqConvAiry2}
\lim_{t\to\infty}h^{\rm resc}(u,t)=\kappa_v {\cal A}_2(u/\kappa_h),
\end{equation}
where $\kappa_v$ and $\kappa_h$ are some vertical and horizontal scaling coefficients depending only on the macroscopic position $\xi$. The Airy$_2$ process has been discovered in the PNG droplet model by Pr\"ahofer and Spohn~\cite{PS02}, and later shown to occur in several other models~\cite{Jo03,Jo03b,Jo02b,SI03,SI07,FS03,BFS07}.\\[0.5em]
\emph{(b) Straight macroscopic limit shape} ($h_{\rm ma}''(\xi)=0$ in a neighborhood of $\xi$). The limit process is the Airy$_1$ process, ${\cal A}_1$,
\begin{equation}\label{eqConvAiry1}
\lim_{t\to\infty}h^{\rm resc}(u,t)=\kappa_v {\cal A}_1(u/\kappa_h).
\end{equation}
The Airy$_1$ process was discovered in the totally asymmetric simple exclusion process (TASEP) by Sasamoto~\cite{Sas05}, see also~\cite{BFPS06,BFP06}. Recently, we proved (\ref{eqConvAiry1}) for the PNG model with flat growth too~\cite{BFS07b}. All these convergence are in the sense of finite-dimensional distributions.

For a precise a definition of the Airy$_1$ and Airy$_2$ processes, we refer to the original papers~\cite{PS02,Sas05,BFPS06} and the review~\cite{Fer07}, in which also known properties are summarized, e.g., the one-point distribution given by the Tracy-Widom distributions of random matrix~\cite{TW02}. It can also happen that curved part of the surface becomes straight (smoothly). Then, the surface will be described by the transition process studied in~\cite{BFS07}. The last typical case is stationary growth, see~\cite{PS02b,BR00,Pra03}, to which we come back at the end of the paper.

\subsubsection*{Limit processes at different times: along space-like paths}
The above results concern the description of the interface at a given large time $t$. However, they do not say anything about the correlations of the surface height at \emph{different times}. First in~\cite{BO04} for the PNG droplet and more recently in~\cite{BFS07b} for flat PNG, results along \emph{space-like} paths have been obtained. Let us remind what it is meant with space-like paths. A path $t=\pi(x)$ in space-time is then called space-like if $|\pi'(x)|\leq 1$ for all $x$.

\begin{figure}[t!]
\begin{center}
\psfrag{x}{$x$}
\psfrag{t}{$t$}
\includegraphics[height=4cm]{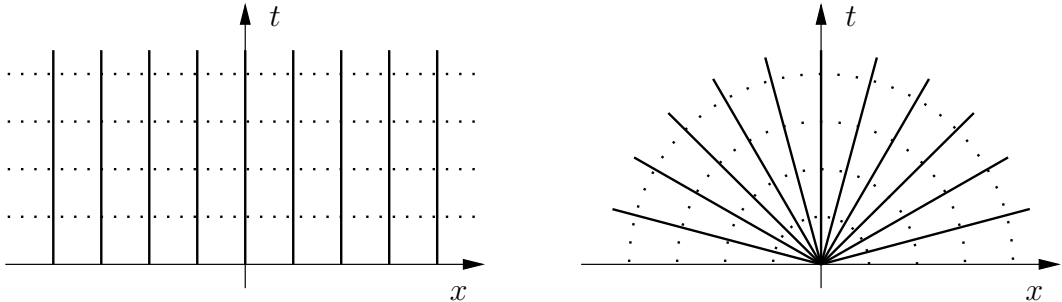}
\caption{Fibration of space-time for (left) flat PNG and (right) PNG droplet. The bold lines indicate the slow directions.}
\label{FigFibration}
\end{center}
\end{figure}

Suppose that we want to analyze the surface of the PNG droplet in a neighborhood of the space-time position $(\xi T,T)$, for some $\xi\in (-1,1)$ and $T\gg 1$. For simplicity, we choose $\pi(\xi)=1$ and set
\begin{equation}\label{eqDropletParam}
x(u)=\xi T+u T^{2/3},\quad t(u)=T\pi(x(u)/T).
\end{equation}
Borodin and Olshanski prove~\cite{BO04} that for space-like paths
\begin{equation}\label{eqSpaceLikeAiry2}
\lim_{T\to\infty}\frac{h(x(u),t(u))-t(u) h_{\rm ma}(x(u)/t(u))}{t(u)^{1/3}}=\kappa_v{\cal A}_2(u (1-\xi\pi'(\xi))/\kappa_h)
\end{equation}
with $\kappa_v=(1-\xi^2)^{1/6}$ and $\kappa_h=(1-\xi^2)^{5/6}$.
In particular, the correlations between $h(x(0),t(0))$ and $h(x(u),t(u))$ with $\pi'(\xi)=0$ are identical to the correlations between $h(x(0),t(0))$ and $h(x(u_\pi),t(u_\pi))$ if \mbox{$u_\pi(1-\xi\pi'(\xi))=u$}. Geometrically, these points form a line with constant slope in the $(x,t)$ diagram (these are the bold lines in Figure~\ref{FigFibration} (right)).

For the flat PNG, the parametrization is slightly simpler because of translation-invariance:
\begin{equation}\label{eqFlatParam}
x(u)=u T^{2/3},\quad t(u)=T \pi(x(u)/T)
\end{equation}
with $\pi(0)=1$. In~\cite{BFS07b} we prove that for space-like paths $\pi$,
\begin{equation}\label{eqSpaceLikeAiry1}
\lim_{T\to\infty}\frac{h(x(u),t(u))-t(u) h_{\rm ma}(x(u)/t(u))}{t(u)^{1/3}}=2^{1/3}{\cal A}_1(u/2^{2/3}).
\end{equation}

\subsubsection*{Slow space-time curves with decorrelation exponent $\tau=1$}
At first sight surprisingly, the scaling coefficients in (\ref{eqSpaceLikeAiry1}) \emph{do not depend on the slope of the space-like path}. Similarly, in (\ref{eqSpaceLikeAiry2}) the correlations are the same if we follow lines with constant angle. Why? The reason is that the space-time is fibred in a non-trivial way: along certain directions the decorrelation is much more slow, on a $T^\tau$ scale for some scaling exponent $\tau>2/3$.

\begin{center}
\textbf{Main result.} By considering the PNG model we show that $\tau=1$.
\end{center}

For the flat PNG, the particular curves are parallel to the time axis, while for the PNG droplet are the one keeping a constant angle. In general, these curves are such that the gradient of the surface is constant: the \emph{characteristics curves} of the following PDE. Let $\textbf{u}=\nabla h$ be the slope of the surface. Then, on a macroscopic scale, $\partial h/\partial t=v^{\rm eff}(\textbf{u})$, where $v^{\rm eff}$ is the effective speed including the noise effect in (\ref{eqKPZequation}). Then, taking the spatial derivative, we get
\begin{equation}
\frac{\partial \textbf{u}}{\partial t}+a(\textbf{u})\frac{\partial \textbf{u}}{\partial x}=0,\quad a(\textbf{u})=-\frac{\partial v^{\rm eff}(\textbf{u})}{\partial \textbf{u}}.
\end{equation}
The characteristics of this equation are the trajectories satisfying $\partial x/\partial t=a(\textbf{u})$ and $\partial \textbf{u}/\partial t=0$.

For the polynuclear growth model, one has $v^{\rm eff}(\textbf{u})=\sqrt{4+\textbf{u}^2}$~\cite{Pra03}, from which one verifies that the slow directions of Figure~\ref{FigFibration} are indeed the characteristics. Why are precisely the characteristics curves to have slow decorrelations? Because the information flows along them, so on a macroscopic scale, the fluctuations of the height is influenced by the randomness occurring around the characteristics.

In general, since characteristics can cross~\cite{CM98} and one has the phenomena of shock formation. In that case, the derivative of the limit shape has a discontinuity. As said in the introduction, in this paper we focus away from this situation. In terms of the slope $\textbf{u}$, we consider cases without shocks, i.e., either with constant $\textbf{u}$ or in the rarefaction regions.

We believe that the same phenomena occurs in higher dimensions too. In a recent work with A.~Borodin~\cite{BF08} on a model in the anisotropic KPZ class in $2+1$ dimensions a similar phenomena of slow curves seems to occur.

\subsubsection*{Consequences}
A first consequence is that, as soon as we do not follow the characteristics, (\ref{eqSpaceLikeAiry1}) and (\ref{eqSpaceLikeAiry2}) hold \emph{without the space-like constraint}. To illustrate it, let us state the precise result for joint distributions in the PNG droplet, which is proven in Section~\ref{SectProofDroplet}.
\begin{thm}\label{ThmPNGDroplet}
Consider the parametrization (\ref{eqDropletParam}) but this time $\pi$ is any space-time path with $\pi(\xi)=1$ and $\pi'(\xi)\neq 1/\xi$, i.e., we exclude the case of the path $\pi$ following the characteristics. Then, in the sense of finite-dimensional distributions,
\begin{equation}
\lim_{T\to\infty}\frac{h(x(u),t(u))-t(u) h_{\rm ma}(x(u)/t(u))}{t(u)^{1/3}}=\kappa_v{\cal A}_2(u (1-\xi\pi'(\xi))/\kappa_h)
\end{equation}
with $\kappa_v=(1-\xi^2)^{1/6}$ and $\kappa_h=(1-\xi^2)^{5/6}$.
\end{thm}

The analogue for the flat PNG is that (\ref{eqSpaceLikeAiry1}) holds for any finite $\pi'$, since in that case the characteristics corresponds to $\pi'=\infty$. Therefore, to get a different behavior from the Airy processes for the height-height correlation, one has to rescale the component of the space-time path along characteristics with the scaling exponent $\tau=1$ and the other component using the scaling exponent $2/3$. For example, for the flat PNG, one has consider the scaling
\begin{equation}\label{eqScaling}
x(u)=u T^{2/3},\quad t(v)=T(1+v).
\end{equation}
Then, the observable to be analyzed is the two-parameter limit process of the height function rescaled as
\begin{equation}\label{eqScaling2}
h_T^{\rm resc}(u,v):=\frac{h(x(u),t(v))-t(v)h_{\rm ma}(x(u)/t(v))}{t(v)^{1/3}}.
\end{equation}
The description of this limit process remains an open problem.

\subsubsection*{Outline}
The rest of the paper is organized as follows. First we give a heuristic argument for $\tau$ can not be strictly less than one. Then we prove it for the PNG droplet (Theorem~\ref{ThmDroplet}). This needs also a technical control of the maximum (Corollary~\ref{corollary}) for which tightness is needed.
For the flat PNG geometry a similar result will hold (Claim~\ref{PropFlat}). In these two cases, one actually sees that the heuristic argument is the right interpretation. Finally, in Proposition~\ref{PropStatPNG}, we show the same result for stationary growth, which turns out to be much simpler, but in the proof there is no trace of the mechanism.

\subsubsection*{Acknowledgments}
The first time I though about this problem was after a short discussion with H.~Spohn and B.~Vir{\'a}g at the Random Matrix Conference during the Hausdorff Research Institute for Mathematics, Bonn (2008). I'm grateful to H.~Spohn and A.~Borodin for helpful discussions.

\section{Results and proofs}
In this section we prove that for the PNG model, for any $\tau<1$, if we take two points along the special directions at time $T$ and time $T+T^\tau$, then they are trivially correlated on the fluctuation scale $T^{1/3}$. But first let us present a heuristic argument, which then it is transformed into a proof for the PNG model.

\subsection{Heuristic argument}
\begin{figure}[t!]
\begin{center}
\psfrag{x}[l]{$x$}
\psfrag{t}[l]{$t$}
\psfrag{T1}[br]{$T$}
\psfrag{T2}[br]{$T+T^\tau$}
\psfrag{Zoom}[c]{Zoom}
\psfrag{T23}[bc]{$T^{2/3}$}
\psfrag{Tmu}[bc]{$T^\nu$}
\psfrag{A}[b]{$A$}
\psfrag{B}[bc]{$B$}
\includegraphics[height=5cm]{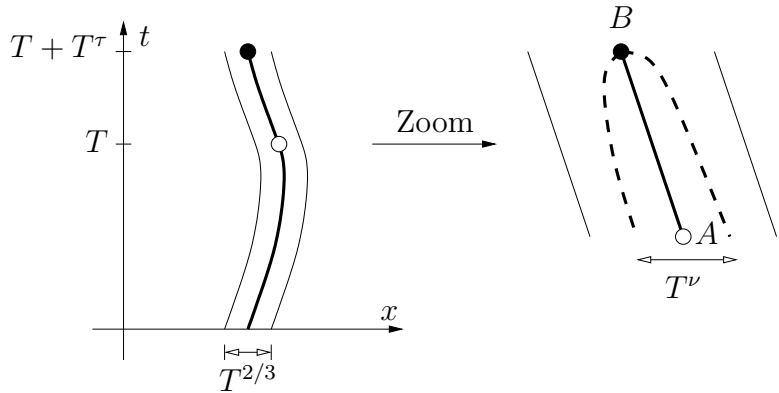}
\caption{The white and black dots are two space-time positions lying on the same characteristic curve (the bold curve). The height fluctuations along this curve are correlated with events in space-time inside a small strip of order $T^{2/3}$. In the zoom, it is indicated that the correlations of the events after time $T$ lies in a smaller region, of width at most $T^\nu$, with $\nu<2/3$ for $\tau<1$.}
\label{figheuristics}
\end{center}
\end{figure}
On a macroscopic level, the information flows along the characteristic curves. Consider two positions on one of such curves, say $A$ at time $T$ and $B$ at time $T+T^\tau$ (the white and black dots in Figure~\ref{figheuristics}). We have the following facts:
\begin{itemize}
\item[(1)] The correlation exponent is $2/3$, therefore the fluctuations of the heights at $A$ and $B$ are correlated with the randomness occurring in a strip of order $T^{2/3}$ around the characteristics. However, if $\tau<1$, the events occurring in the time interval $[T,T+T^\tau]$ will be relevant only on a smaller region, of order $T^\nu$ with $\nu=2\tau/3$.
\item[(2)] Again, because the correlation exponent is $2/3$, the fluctuations of the height function at time $T$ in a $T^\nu$-neighborhood of $A$ differ from the fluctuations of the height at $A$ by ${\cal O}(T^{\nu/2})=o(T^{1/3})$.
\item[(3)] The fluctuation exponent is $1/3$, meaning that during a time interval $T$, fluctuations are created on a $T^{1/3}$ scale.
\end{itemize}
By (1), the height at $B$ is the sum of the height at some point in a $T^\nu$-neighborhood of $A$ plus the contribution during the remaining $T^\tau$ time interval. By (2), the fluctuations of the first term is equal to the fluctuations of the height at $A$ up to an $o(T^{1/3})$ term. By (3), the fluctuations in the second term are only of order $T^{\tau/3}=o(T^{1/3})$. Therefore, on the natural scale of fluctuations, $T^{1/3}$, the height of $A$ and $B$ are trivially correlated in the $T\to\infty$ limit (i.e., the difference is only a deterministic factor).

Remark that, like in the PNG model defined below, often the characteristics curves are straight lines. However, for generic time-dependent dynamics, this is not the case.

\subsection{PNG and directed polymers}
\begin{figure}[t!]
\begin{center}
\psfrag{u}[l]{$u$}
\psfrag{v}[l]{$v$}
\psfrag{x}[l]{$x$}
\psfrag{t}[l]{$t$}
\psfrag{h(x,t)}[l]{$h(x,t)$}
\psfrag{(0,0)}[r]{$(0,0)$}
\psfrag{(t1,t1)}[l]{$(x+t,t-x)$}
\includegraphics[height=5.5cm]{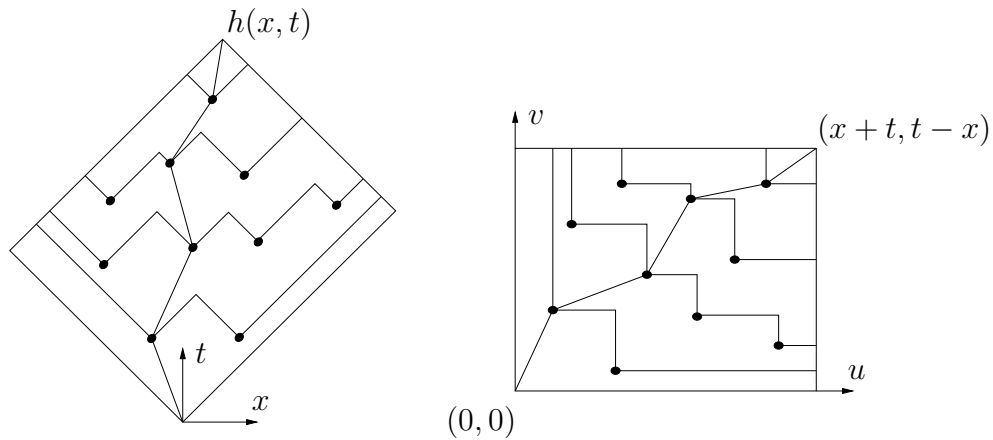}\caption{Height and directed polymers for the droplet geometry.}\label{figDP1}
\end{center}
\end{figure}

In the proof, we use the picture of directed percolation on Poisson points, whose connection to the PNG model is the following. Let us start with the PNG droplet case. The nucleations in the PNG model form a Poisson point process in space-time with intensity $\rho(x,t)=2\Id(|x|\leq t)$. We turn by $\pi/4$ the point of view as in Figure~\ref{figDP1}. To avoid annoying $\sqrt{2}$ factors, we then zoom out by a linear factor $\sqrt{2}$, so that the density of Poisson points in the directed percolation picture is $1$.

Consider all the directed paths consisting of a sequences of Poisson points, where the points are joined by segments of slope in $[0,\infty]$. The length of a path is the number of Poisson points visited. Then, for a point $(u,v)\in\R_+^2$, let $L((0,0),(u,v))$ be the length of the longest path from $(0,0)$ to $(u,v)$. We call \emph{maximizers} the paths of longest length. As noticed in~\cite{PS00}, the height function of the PNG droplet at position $(x,t)$, $h(x,t)$, can be expressed by $L$, namely
\begin{equation}\label{eqDP1}
h(x,t)=L((0,0),(x+t,t-x)).
\end{equation}

For the flat PNG, the only difference is $\rho(x,t)=2\Id(t\geq 0)$, which implies that the height function is given by (\ref{eqDP1}) but for $L$ the optimization is made line-to-point: from $\{u+v=0\}$ to $(x+t,t-x)$. More details can be found in the literature, for example in Section 3 of~\cite{FP05}.

\subsection{PNG droplet}\label{SectProofDroplet}
Denote by $\mathbf{h}(x,t)=h(x,t)-t h_{\rm ma}(x/t)$ the difference between the height function and the approximate position from the limit shape.
\begin{thm}[PNG droplet]\label{ThmDroplet} Consider a fixed fixed $\xi\in (-1,1)$. For any $\tau<1$, there exists a $\beta<1/3$ such that
\begin{equation}
\lim_{T\to\infty}\Pb(|\mathbf{h}(\xi T,T)-\mathbf{h}(\xi (T+T^\tau),T+T^\tau)|\leq T^\beta)=1.
\end{equation}
\end{thm}

To transform the heuristic argument into the proof of Theorem~\ref{ThmDroplet}, we use results of~\cite{FS03} which control the geometric distribution of the maximizers close to the end points. The results in~\cite{FS03} use ideas from in~\cite{Jo00} about transversal fluctuations. Then, it will only remain one technical difficulty: the convergence to the Airy$_2$ process of~\cite{PS02,BO04} are in the sense of finite-dimensional distributions. This is not enough to conclude that the maximum converges: a tightness result is needed.

\begin{proofOF}{Theorem~\ref{ThmDroplet}}
First, notice that it is enough to consider the case $\xi=0$. Indeed, if we consider a point $(\xi T,T)$ and apply the mapping
\begin{equation}\label{eqMapXiZero}
\Phi:(x,t)\mapsto (1-\xi^2)^{-1/2}(x-t\xi,t-\xi x),
\end{equation}
we recover the situation with $\xi=0$ with time given by $\sqrt{1-\xi^2}T$ instead of $T$. This mapping preserves the Poisson measure on nucleations and, in the percolation picture directed paths remain directed. Therefore, w.l.o.g, we consider the $\xi=0$ situation.

\begin{figure}[t!]
\begin{center}
\psfrag{A}[cb]{$A$}
\psfrag{B}[cb]{$B$}
\psfrag{H}[cb]{$\cal H$}
\psfrag{S}[l]{$S(\nu)$}
\psfrag{O}[c]{$O$}
\psfrag{x}[cb]{$x$}
\psfrag{t}[l]{$t$}
\psfrag{tnu}[c]{$T^\nu$}
\psfrag{talpha}[r]{$T^\tau$}
\psfrag{Talpha}[l]{$T^\tau$}
\psfrag{T}[l]{$T$}
\includegraphics[height=5cm]{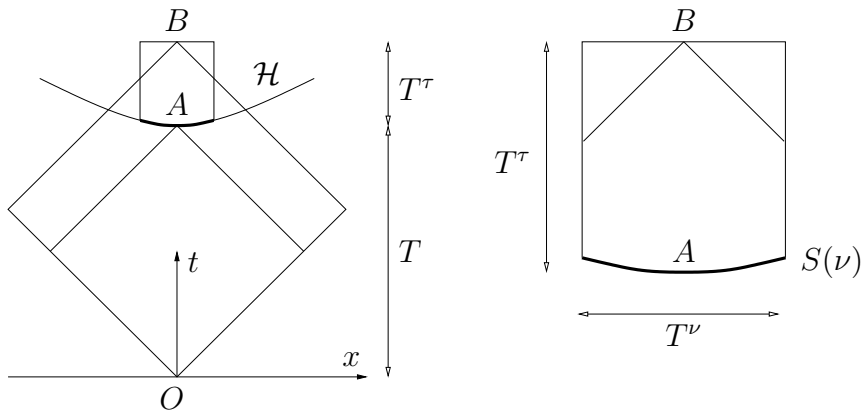}
\caption{Geometric construction. The coordinates of the points are \mbox{$O=(0,0)$}, $A=(0,T)$, and $B=(0,T+T^\tau)$. }
\label{FigGeomConstruction}
\end{center}
\end{figure}

As illustrated in Figure~\ref{FigGeomConstruction}, denote $O=(0,0)$, $A=(0,T)$, and \mbox{$B=(0,T+T^\tau)$}. Moreover, let $C(\tau,\nu)$ be the cylinder of axis $AB$ and width $w=2T^\nu$, and $\partial C(\tau,\nu)$ be the border border of the cylinder without lids. We consider the following event
\begin{multline}
D=\{\textrm{Poisson points configurations such that the maximizers}\\ \textrm{from }O\textrm{ to }B\textrm{ do not cross }\partial C(\tau,\nu)\}.
\end{multline}
This quantity was studied in the proof of Theorem 1(ii) of~\cite{FS03}. To be more precise, one uses the bound on $\Pb(\Omega\setminus D)$ in the equation just after (20) with the following improvements. The $\e$ is the one of Lemma 2 and it is actually $t$-dependent by using (11). The bound on $\Pb(F)$ is in the equation after (22), which is also $t$-dependent by using the bound of~\cite{BDJ99} reported in (8).
In the case $\tau<1$ and setting $\nu=\tfrac23-\tfrac{1-\tau}{6}$, we obtain
\begin{equation}\label{eq2.4}
\Pb(D)\geq 1-c_1 \exp\left(-c_2 T^{(1-\tau)/12}\right)
\end{equation}
for some constant $c_1,c_2>0$ independent of $t$.

Consider the hyperbole $\cal H$, along which the height is macroscopically constant ($\simeq 2T$):
\begin{equation}\label{eqhyperbole}
{\cal H}=\{(x,t)=(r(\theta)\sin(\theta),r(\theta)\cos(\theta)), r(\theta)=T\sqrt{\cos(2\theta)},\theta\in (-\pi/4,\pi/4)\}
\end{equation}
and let $S(\nu)$ be the piece of $\cal H$ lying inside the cylinder $C(\tau,\nu)$. $\theta$ in (\ref{eqhyperbole}) is the angle between the $x=0$ axis and the segment from $O$ to the point in $\cal H$.

The bound (\ref{eq2.4}) tell us that, with probability going to one as $T\to\infty$, the maximizers of $L(O,B)$ crosses $S(\nu)$. Therefore we need to concentrate on those situations only, for which we have
\begin{equation}\label{eqIneq}
L(O,A)+L(A,B)\leq L(O,B)\leq L(O,S)+L(S,B).
\end{equation}
We rewrite it as
\begin{equation}\label{eq2.7}
L(A,B)-2T^\tau\leq L(O,B)-L(O,A)-2T^\tau\leq L(O,S)-L(O,A)+L(S,B)-2T^\tau.
\end{equation}
Recall that $L(O,A)\simeq 2T$, $L(O,B)\simeq 2(T+T^\tau)$, and $L(A,B)\simeq 2 T^\tau$. Thus, we have to show that both right and left hand side are (in absolute value) smaller than $T^\beta$ with probability $1$ as $T\to\infty$, for some $\beta<1/3$.

From Lemma 7.1 of~\cite{BDJ99}, setting $\beta=(1+\tau)/6$, we have
\begin{equation}\label{eq2.9}
\Pb(|L(A,B)-2T^\tau|\geq T^\beta)\leq c_1\exp(-c_2 T^{(1-\tau)/4}).
\end{equation}
Moreover, since $L(A,B)\leq L(S,B)\leq L((x,t=T),B)$ and both terms are $2T^\tau + {\cal O}(T^{\tau/3})$ ~\cite{BR00,PS00}, by Markov inequality it follows
\begin{equation}
\Pb(|L(S,B)-2T^\tau|\geq T^\beta)\leq {\cal O}(1/T^{(1-\tau)/6})\to 0 \quad \textrm{ as }T\to\infty.
\end{equation}
A better bound of the type (\ref{eq2.9}) can be also obtained using the results of~\cite{BR99b}, but it is not needed for our purposes.

Finally, consider the points
\begin{equation}\label{eq2.30}
X(u)=(x(u),t(u))\textrm{ on }\cal H
\end{equation}
associated with the angle $\theta=u T^{-1/3}$. Then, $u\mapsto (L(O,X(u))-2T)/T^{1/3}$ converges, in the sense of finite-dimensional distributions, to the Airy$_2$ process ${\cal A}_2(u)$~\cite{FPS03,BO04}. The Airy$_2$ process is a.s.\ continuous and the points in $S(\nu)$ corresponds to $|u|\leq T^{\nu-2/3}=1/T^{(1-\tau)/6}$. The control of the maximum requires a bit more than finite-dimensional distributions, which is the content of the next part of the paper leading to Corollary~\ref{corollary}: there exists a $\beta<1/3$ such that $\Pb(|L(O,S)-L(O,A)|\geq T^\beta)\to 0$ as $T\to\infty$. This end the proof.
\end{proofOF}

With this Theorem we can then prove Theorem~\ref{ThmPNGDroplet}.
\begin{proofOF}{Theorem~\ref{ThmPNGDroplet}}
Using the transformation (\ref{eqMapXiZero}), we consider w.l.o.g.\ the $\xi=0$ case. As in Theorem~\ref{ThmDroplet}, denote $\mathbf{h}(x,t)=h(x,t)-t h_{\rm ma}(x/t)$ and consider $m$ points parametrized by space-time positions $x_i=u_i T^{2/3}$ and $t_i=T\pi(u_i T^{-1/3})$. Then we want to determine the large $T$ limit of
\begin{equation}\label{eq2.12}
\Pb\left(\bigcap_{k=1}^m \{\mathbf{h}(x_k,t_k)\leq s_k T^{1/3}\}\right)
=\Pb\left(\bigcap_{k=1}^m \{\mathbf{h}(x_k,T)\leq s_k T^{1/3}+ \Xi_k\}\right)
\end{equation}
where $\Xi_k=\mathbf{h}(x_k,T)-\mathbf{h}(x_k,t_k)$. Denote the events $Q_k=\{|\Xi_k|\leq T^\beta\}$, and $A_k=\{\mathbf{h}(x_k,T)\leq s_k T^{1/3}+ \Xi_k\}$. Then we can rewrite (\ref{eq2.12}) as
\begin{equation}\label{eq2.12b}
(\ref{eq2.12})=\Pb\left(\bigcap_{k=1}^m \left(\left(A_k\cap Q_k\right)\cup\left(A_k\cap Q_k^c\right)\right)\right)
=\Pb\left(\left(\bigcap_{k=1}^m (A_k\cap Q_k)\right)\cup R\right)
\end{equation}
where in $R$ are terms with at least one $A_j\cap Q_j^c$. Then, using inclusion-exclusion principle, we get
\begin{equation}\label{eq2.12c}
(\ref{eq2.12b})=\Pb\left(\bigcap_{k=1}^m (A_k\cap Q_k)\right)+\Pb(R)-\Pb\left(R\cap \bigcap_{k=1}^m (A_k\cap Q_k)\right).
\end{equation}
By Theorem~\ref{ThmDroplet}, there exists a $\beta<1/3$ such that $\Pb(Q_j)\to 1$ as $T\to\infty$, i.e., $\Pb(Q_j^c)\to 0$. Since the last two terms contain at least one $Q_j^c$, they go to zero as $T\to\infty$. For the first term in the r.h.s.\ of (\ref{eq2.12c}), because $\beta<1/3$, we have
\begin{multline}
\lim_{T\to\infty}\Pb\left(\bigcap_{k=1}^m ((\mathbf{h}(x_k,T)\leq s_k T^{1/3}+ \Xi_k)\cap \{|\Xi_k|\leq T^\beta\})\right) \\
=\lim_{T\to\infty}\Pb\left(\bigcap_{k=1}^m (\mathbf{h}(x_k,T)\leq s_k T^{1/3})\right)=\Pb\left(\bigcap_{k=1}^m ({\cal A}_2(u_k)\leq s_k)\right),
\end{multline}
where the last equality is the fixed time result of~\cite{PS02}.
\end{proofOF}

\subsubsection*{Multilayer PNG droplet}
\begin{figure}[t!]
\begin{center}
\psfrag{x=-t}[c]{$x=-T$}
\psfrag{x=t}[c]{$x=T$}
\psfrag{h0}[l]{$h_0$}
\psfrag{h1}[l]{$h_{-1}$}
\psfrag{h2}[l]{$h_{-2}$}
\includegraphics[height=5.5cm]{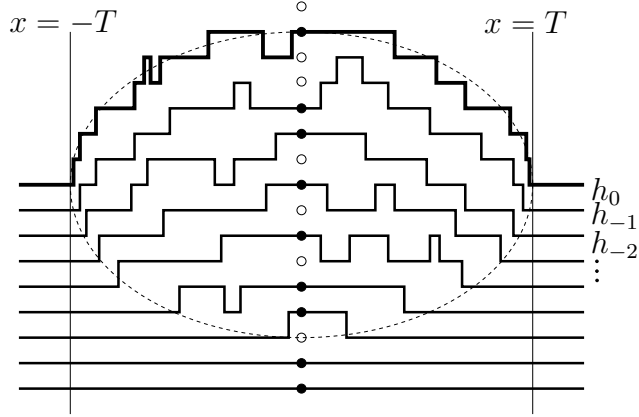}
\caption{Illustration of the multilayer version of the polynuclear growth model in the PNG droplet geometry at time $t=T$.}
\label{FigMultilayer}
\end{center}
\end{figure}
To obtain the result of Corollary~\ref{corollary}, we use the multilayer extension of the polynuclear growth model introduced in~\cite{PS02}. Let us just recall it briefly. Instead of a single height function $h(x,t)$, we have a set of height functions $\{h_\ell(x,t),\ell\leq 0\}$, where $h_0(x,t)$ coincides with the original $h(x,t)$. Initially, $h_\ell(x,0)=\ell$, for all $x\in\R$. A nucleation at level $\ell$, for $\ell\leq -1$, occurs whenever there is a merging at level $\ell+1$ and the nucleation is at the same position of the merging. Moreover, the deterministic part of the dynamics for the PNG applies to the new height functions as well (see (i) at page~\pageref{PNGdynamics}). By construction the lines do not intersect (see Figure~\ref{FigMultilayer} for an illustration).

One then defines an extended points process $\eta$ by
\begin{equation}
\eta((x,t),j)=\left\{\begin{array}{ll}
1,& h_\ell(x,t)=j\textrm{ for some }\ell\leq 0,\\ 0,&\textrm{otherwise}
\end{array} \right.
\end{equation}
Along the hyperbole $\cal H$ defined in (\ref{eqhyperbole}) the point process $\eta$ is \emph{stationary}, if we choose as parametrization the angle $\theta$ with the origin~\cite{BO04}. For any point $(x_i,t_i)\in {\cal H}$, we denote by $\theta_i$ the corresponding angle. As shown in~\cite{BO04,FPS03}, $\eta$ is an extended determinantal point process, i.e., it has determinatal correlation functions: $\rho^{(n)}(\theta_1,j_1;\ldots;\theta_m,j_m)=\det[K_T(\theta_k,j_k;\theta_l,j_l)]_{1\leq k,l\leq m}$ with $K_T$ the extended Bessel kernel. Define the operator $H$ on $\ell^2(\Z)$ by
\begin{equation}
(H \psi)(j)=-\psi(j+1)-\psi(j-1)+2\psi(j)+\frac{j}{T}\psi(j)
\end{equation}
and the one-time Bessel kernel by
\begin{equation}\label{eq2.13}
B_T(i,j)=\sum_{\ell\geq 0}J_{i+\ell}(2T)J_{j+\ell}(2T),
\end{equation}
with $J_n(2T)=\frac{1}{2\pi\I}\oint_{|z|=1}\dx z e^{T(z-1/z)} z^{-n-1}$ the standard Bessel functions~\cite{AS84}. Then, the stationary extended Bessel kernel is given by
\begin{equation}
K_T(\theta_1,j_1;\theta_2,j_2)=\left\{\begin{array}{ll}
(e^{-T\theta_1 H} B_T e^{T\theta_2 H})(j_2,j_1),& \theta_2\geq \theta_1,\\[0.5em]
(e^{-T\theta_1 H} (B_T-\Id) e^{T\theta_2 H})(j_2,j_1),&\theta_2<\theta_1.
\end{array}\right.
\end{equation}
\noindent\emph{Rescaled process}. Recall from (\ref{eq2.30}) that for the rescaled process on $\cal H$ the angle is $\theta=u/T^{1/3}$, which is thus given by
\begin{equation}
\eta_T^{\rm resc}(u_i,s_i)=T^{1/3} \eta_T(u_i T^{-1/3},[2T+s_i T^{1/3}]).
\end{equation}
Then, denote the rescaled height function by
\begin{equation}
h_T^{\rm resc}(u)=\frac{h(x(u),t(u))-T h_{\rm ma}(x(u)/t(u))}{T^{1/3}}
\end{equation}
with $(x(u),t(u))$ the coordinate on $\cal H$ with angle $\theta=u/T^{1/3}$. Also, for a given function $f$, we define
\begin{equation}
f_T(x)=f((x-2T)/T^{1/3}),
\end{equation}
so that
\begin{equation}\label{eq2.18}
\eta_T^{\rm resc}(u,f)=\frac{1}{T^{1/3}}\sum_{s\in (\Z-2T)/T^{1/3}} f(s)\eta_T^{\rm resc}(u,s) = \sum_{j\in\Z} f_T(j)\eta_T(u T^{-1/3},j).
\end{equation}

After this setting, we can go to the proof. The strategy is almost identical as in Section 5 of~\cite{Jo03b}. The only difference, is the derivation of the first lemma (Lemma 5.1 in~\cite{Jo03b}), which is carried out as in Appendix A of~\cite{PS02}, exploiting the stationarity and the semi-group structure in the kernel.

\begin{lem}\label{lemma5.1}
Assume that $f$ is a $C^\infty$ function and there exists a constant $M$ such that $f(x)=0$ if $x\leq -M$ and $f(x)=1$ if $x\geq M$. Then, there is a constant $C(f)$ such that
\begin{equation}\label{eqLemma5.1}
\E\left((\eta_T^{\rm resc}(u_i,f)-\eta_T^{\rm resc}(u_j,f))^4\right)\leq C(f) \left(|u_i-u_j|^2+|u_i-u_j|/T^{2/3}\right)
\end{equation}
for $-1\leq u_i,u_j\leq 1$ and for all $T\geq T_0$, with $T_0$ a fixed large enough constant.
\end{lem}
This lemma is the corresponding of Lemma 5.1 in~\cite{Jo03b} if one takes into account that in~\cite{Jo03b} time is discrete, which means that $|u_i-u_j|\geq 1/T^{2/3}$.
\begin{proofOF}{Lemma~\ref{lemma5.1}}
First of all, using (\ref{eq2.18}) we get for the l.h.s.~of~(\ref{eqLemma5.1}),
\begin{multline}
\E\left((\eta_T^{\rm resc}(u_i,f)-\eta_T^{\rm resc}(u_j,f))^4\right) \\
=\sum_{n=0}^4 \binom4n (-1)^n \sum_{i_1,\ldots,i_4\in\Z} f_T(i_1)\cdots f_T(i_4)
\E\left(\eta_T(u_1^{(n)}T^{-1/3},i_1)\cdots\eta_T(u_4^{(n)}T^{-1/3},i_4)\right)
\end{multline}
where $u_k^{(n)}=u_i$ if $k>n$ and $u_k^{(n)}=u_j$ if $k\leq n$. The last term in the expected value is just the 4-point correlation functions, which is given by a $4\times 4$ determinant.

Now, the reason of considering the height function on the line $\cal H$, is that we have a stationary determinantal point process with kernel $K_T$. $K_T$ is the evolution of one-time kernel $B_T$ by the operator $H$. This is exactly the same setting of Appendix A of~\cite{PS02}. One can copy step by step the proof of Lemma A.1 in~\cite{PS02} except the last one, which is the only one which uses the particular structure of the operator $H$. The result, is a bound $C(f)|u_i-u_j|^2$ plus the following term (see (A.16) of~\cite{PS02})
\begin{equation}\label{eq2.21}
(u_i-u_j) T^{2/3} \Tr\left(B_T F_T\right)
\end{equation}
where $F_T$ is given by
\begin{equation}
F_T:=[H,f_T] f_T^3-f_T^3 [H,f_T]-3 f_T [H,f_T] f_T^2+3f_T^2[H,f_T]f_T.
\end{equation}
It is not difficult to get an expression for $F_T$ (just use a matrix representation for $H$ and $f_T$). We obtain
\begin{equation}
F_T(i,j)=\left\{\begin{array}{ll}
(f_T(i)-f_T(j))^4, & |i-j|=1,\\
0,&\textrm{ otherwise.}
\end{array}\right.
\end{equation}
Therefore,
\begin{equation}\label{eq2.24}
\Tr\left(B_T F_T\right) = \sum_{i,j\in\Z} B_T(i,j) F_T(j,i)=2\sum_{i\in\Z} B_T(i,i+1) (f_T(i+1)-f_T(i))^4,
\end{equation}
where we also used the symmetry of the kernel $B_T$, given in (\ref{eq2.13}). Recall that we have $f_T(i+1)-f_T(i)=0$ if $i\not\in [2T-M T^{1/3},2T+M T^{1/3}]\cap\Z$. Also, $|f_T(i+1)-f_T(i)|\simeq \|f'\|_\infty /T^{1/3}$. Thus,
\begin{equation}
|(\ref{eq2.24})| \leq \cte \|f'\|_\infty^4 T^{-4/3} \sum_{i\in [2T-M T^{1/3},2T+M T^{1/3}]\cap\Z} |B_T(i,i+1)|.
\end{equation}
It remains to see that the last sum is bounded by a constant. The sum is bounded by
\begin{equation}\label{eq2.25}
\sum_{s\in [-M,M]\cap \Z/T^{1/3}} \sum_{\lambda\in \N/T^{1/3}} |J_{[2T+(s+\lambda) T^{1/3}]}(2T)| |J_{[2T+(s+\lambda) T^{1/3}+1]}(2T)|.
\end{equation}
By using the uniform bounds (for $T\geq T_0$, $T_0$ fixed) reported in Lemma A.1 of~\cite{Fer04} (the first bound is due to Landau~\cite{Lan00}):
\begin{eqnarray}\label{eq2.26}
&|T^{1/3}J_{[2T+s T^{1/3}]}(2T)|\leq \cte, &\textrm{ for all }s, \nonumber \\
&|T^{1/3}J_{[2T+s T^{1/3}]}(2T)|\leq \cte e^{-s/2},& \textrm{ for }s\geq 0,
\end{eqnarray}
it follows that $|(\ref{eq2.25})|\leq \cte$ with the constant independent of $T$.
Putting all together, we get
\begin{equation}
|(\ref{eq2.21})|\leq \cte |u_i-u_j| T^{-2/3},
\end{equation}
with $\cte$ depending only on $f$. The uniformity of $C(f)$ for $T\geq T_0$ in the term $C(f)|u_i-u_j|^2$ is also verified using the exponential bounds (\ref{eq2.26}).
\end{proofOF}

For a short while, consider only the point process at a discrete set of angles $\theta_i=u_i T^{-1/3}$, with $u_i=i/T^{2/3}$, $i\in \Z\cap[-T^{2/3},T^{2/3}]$.
\begin{thm}\label{theoremWeakConv}
Let $h_T^{\rm resc}(u)$ be defined on $u\in (T^{-2/3} \Z)\cap[-1,1]$ and by linear interpolation. Then, there is a continuous version of ${\cal A}_2(u)$ and
\begin{equation}
h_T^{\rm resc}(u)\to {\cal A}_2(u)
\end{equation}
as $T\to\infty$ in the weak*-topology of probability measures on $C([-1,1])$.
\end{thm}
\begin{proofOF}{Theorem~\ref{theoremWeakConv}}
For $u_i=i/T^{2/3}$ and integer $i$, the bound of Lemma~\ref{lemma5.1} is simply given by $C(f) |u_i-u_j|^2$ because $|u_i-u_j|\geq T^{-2/3}$. This is the analogue of Lemma 5.1 of~\cite{Jo03b}. Moreover, all the work of Section~5.2 of~\cite{Jo03b} is valid here too, without modifications. Since the convergence of finite dimensional distribution was proven in~\cite{BO04}, we have weak*-convergence of $h_T^{\rm resc}$ (it is the corresponding of Theorem 1.2 of~\cite{Jo03b}).
\end{proofOF}

The final step is to use Theorem~\ref{theoremWeakConv} together with a control of the maximum in the intervals between points in $\Z/T^{2/3}$ to get the convergence of the maximum.
\begin{lem}\label{lemmaShortDistances}
For $i\in J=\Z\cap[-T^{2/3},T^{2/3}]$ we have the uniform bound (for $T\geq 1$)
\begin{multline}\label{eq2.28}
\Pb\left(\max_{0\leq x\leq 1/T^{2/3}}h_T^{\rm resc}(u_i+x)-\max\{h_T^{\rm resc}(u_i),h_T^{\rm resc}(u_i+T^{-2/3})\}\geq T^{-1/6}\right)\\ \leq \cte e^{-T^{1/6}},
\end{multline}
from which it follows
\begin{equation}
\Pb\left(\max_{-1\leq x\leq 1}{h_T^{\rm resc}(u_i+x)-\max_{i\in J}h_T^{\rm resc}(u_i)}\geq T^{-1/6}\right)\leq \cte T^{2/3}e^{-T^{1/6}}.
\end{equation}
\end{lem}
\begin{proofOF}{Lemma~\ref{lemmaShortDistances}}
Let ${\cal R}$ be the shaded region  of Figure~\ref{FigSmallDistances}.
\begin{figure}[t!]
\begin{center}
\psfrag{x}[c]{$x$}
\psfrag{t}[l]{$t$}
\psfrag{A}[c]{$A$}
\psfrag{O}[c]{$O$}
\psfrag{X}[cb]{$X_1$}
\psfrag{Y}[cb]{$X_2$}
\includegraphics[height=3cm]{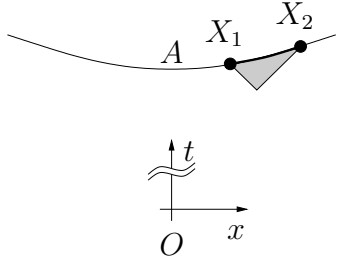}
\caption{Construction for the small distance bound with two points at distance ${\cal O}(1)$ on $\cal H$: $X_1=X(u_i)$ and $X_2=X(u_i+T^{-2/3})$.}
\label{FigSmallDistances}
\end{center}
\end{figure}
Then,
\begin{equation}
\max_{0\leq x\leq 1/T^{2/3}}h_T^{\rm resc}(u_i+x)-\max\{h_T^{\rm resc}(u_i),h_T^{\rm resc}(u_i+T^{-2/3})\}\leq \frac{\textrm{\# Poisson points in }{\cal R}}{T^{1/3}},
\end{equation}
where the factor $T^{-1/3}$ comes from the rescaling in $h_T^{\rm resc}$. The area of $\cal R$ is smaller than $1$. Therefore
\begin{align}
\textrm{l.h.s.~of~}(\ref{eq2.28})&\leq \Pb(\textrm{\# Poisson points in }{\cal R}\geq T^{1/6})\nonumber \\
&\leq \sum_{n\geq T^{1/6}} \frac{\cte}{n!}\leq \cte e^{-T^{1/6}},
\end{align}
for $T\geq 1$.
\end{proofOF}
A direct consequence of Theorem~\ref{theoremWeakConv} and Lemma~\ref{lemmaShortDistances} is resumed in the following corollary.
\begin{cor}\label{corollary}
For any $\delta>0$, there exists a $\beta<1/3$ such that
\begin{equation}
\lim_{T\to\infty} \Pb\left(\max_{u\in [-T^{-\delta},T^{-\delta}]}T^{1/3}(h_T^{\rm resc}(u)-h_T^{\rm resc}(0)) \geq T^{\beta}\right)=0.
\end{equation}
\end{cor}

\subsection{Flat PNG}
Now that we looked in detail the case of the PNG droplet, we just state the corresponding claim for the flat PNG. We present the argument to prove it, but we will not do the tightness argument for that case. Recall that $\mathbf{h}(x,t)=h(x,t)-2t$ for flat PNG.

\begin{claim}[Flat PNG]\label{PropFlat}
For any $\tau<1$, there exists a $\beta<1/3$ such that
\begin{equation}
\lim_{T\to\infty}\Pb(|\mathbf{h}(0,T)-\mathbf{h}(0,T+T^\tau)|\leq T^\beta)=1.
\end{equation}
\end{claim}
The strategy is almost identical to the proof of Theorem~\ref{ThmDroplet} for the PNG droplet. The difference are the following. The line $\cal H$ along which one has a stationary process is just the straight line $(\R,T)$. From the same analysis of the transversal fluctuations in~\cite{Jo00}, we get that for any $\e>0$, the maximizers of $L((\R,0),B)$ start in a $T^{2/3+\e}$-neighborhood of the origin $O$ with probability one in the $T\to\infty$ limit. We can choose for example $\e=(1-\tau)/2$, which implies that the straight line from $(T^{2/3+\e},0)$ to $B$ crosses the lines $(\R,T)$ at a position of order $T^\mu$, with $\mu=\frac23-\frac{1-\tau}{2}$. Then, as we made for (\ref{eq2.4}), from~\cite{FS03} it follows that (a.s.\ as $T\to\infty$) the maximizers in the time interval $[T,T+T^\tau]$ lies inside a cylinder of width $T^\nu$, with $\nu=\frac23-\frac{1-\tau}{6}$. The rest of the argument is the same, but to transform into a mathematical proof one would still need a tightness result, which we do not present in this paper.
\begin{figure}[t!]
\begin{center}
\psfrag{A}[cb]{$A$}
\psfrag{B}[cb]{$B$}
\psfrag{H}[cb]{$\cal H$}
\psfrag{O}[c]{$O$}
\psfrag{x}[cb]{$x$}
\psfrag{t}[l]{$t$}
\psfrag{Tnu}[bc]{$T^\nu$}
\psfrag{talpha}[r]{$T^\tau$}
\psfrag{Talpha}[l]{$T^\tau$}
\psfrag{T}[l]{$T$}
\includegraphics[height=5cm]{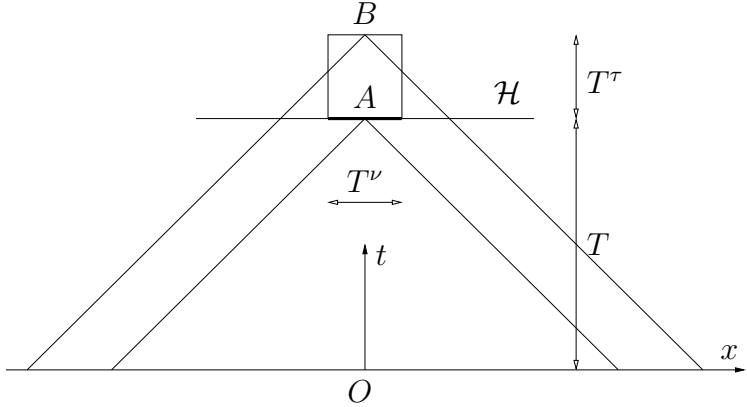}
\caption{Geometric construction for the flat PNG. The coordinates of the points are $O=(0,0)$, $A=(0,T)$, and $B=(0,T+T^\tau)$. The maximization is made from the $(\R,0)$-axis to the points $A$ and $B$.}
\label{FigFlatGeometry}
\end{center}
\end{figure}

\subsection{Stationary PNG}
Finally we consider the stationary growth of the PNG model. The argument is the simplest one, but not very instructive of what actually happens.

The stationary PNG with mean slope zero is described in Chapter 3 of~\cite{Pra03}. The up- and down-steps of the height function are two independent Poisson point processes with density one as shown in Proposition 3.1 of~\cite{Pra03}. Denote by $h^{\rm stat}(x,t)$ the height function at position $x$ and time $t$, where the reference frame is set at time $t=0$ by $h^{\rm stat}(0,0)=0$.

\begin{prop}[Stationary PNG]\label{PropStatPNG}
For any $\tau<1$, there exists a $\beta<1/3$ such that
\begin{equation}
\lim_{T\to\infty}\Pb(|h^{\rm stat}(0,T+T^\tau)-h^{\rm stat}(0,T)-2T^\tau|\leq T^\beta)=1.
\end{equation}
\end{prop}
\begin{proofOF}{Proposition~\ref{PropStatPNG}}
By stationarity,
\begin{equation}\label{eq2.36}
\Pb(|h^{\rm stat}(0,T+T^\tau)-h^{\rm stat}(0,T)-2T^\tau|\leq T^\beta)
=\Pb(|h^{\rm stat}(0,T^\tau)-2T^\tau|\leq T^\beta)
\end{equation}
Set $\beta\in (\tau/3,1/3)$ and use the notation $L=T^\tau$ and $\tilde\beta=\beta/\tau>1/3$. Then,
\begin{equation}\label{eq2.37}
(\ref{eq2.36}) = \Pb(|h^{\rm stat}(0,L)-2L|\leq L^{\tilde\beta}).
\end{equation}
From~\cite{PS00} we know that
\begin{equation}
\lim_{L\to\infty}\Pb(h^{\rm stat}(0,L)-2L\leq s L^{1/3})=F_0(s),
\end{equation}
for some well-defined distribution function $F_0$ (for details on $F_0$ see~\cite{PS00,BR00}). Therefore $\lim_{L\to\infty} (\ref{eq2.37})=1$, because $\tilde\beta>1/3$.
\end{proofOF}

%\newcommand{\bibliodir}[1]{../../Biblio/#1}
%\bibliographystyle{\bibliodir{patplain}}
%\bibliography{\bibliodir{Biblio}}

\end{document}